\documentclass[12pt]{article}
\usepackage[symbol]{footmisc}

\usepackage[left=2cm,top=2.5cm,right=2cm,bottom=2.5cm]{geometry}
\usepackage{amsmath, amssymb}

\usepackage{graphicx}
\usepackage{graphics}
\usepackage{epstopdf}
\usepackage{subfigure}
\usepackage{amsfonts}
\usepackage{sectsty}
\usepackage{sectsty}
\usepackage{hyperref}
\usepackage{cite}
\usepackage{lineno}

\begin{document}
	
	\begin{center}
	\large{\bf{Estimation of Cosmological Parameters, Stability Analysis and Energy Conditions  in Viable Modified Gravity }} \\
	\vspace{5mm}
	\normalsize{Nisha Godani$^1$ and Gauranga C. Samanta$^{2}{}$ }\\
	\normalsize{$^1$Department of Mathematics, Institute of Applied Sciences and Humanities\\ GLA University, Mathura, Uttar Pradesh, India\\
		$^2$Department of Mathematics, BITS Pilani K K Birla Goa Campus, Goa, India}\\
	\normalsize {nishagodani.dei@gmail.com\\gauranga81@gmail.com}
\end{center}

\begin{abstract}

\noindent
In the present paper, we have investigated the Friedmann–Robertson–Walker (FRW) model in viable $f(R,T)$ gravity with $f(R,T)$ function proposed as $f(R,T)=R +\xi T^{1/2}$, where $\xi$ is an arbitrary constant, $R$ is the scalar curvature and $T$ is the trace of stress energy tensor. Defining the scale factor, the field equations are solved numerically and the energy conditions are analyzed. Further, determining Hubble parameter and deceleration parameter, their present values are estimated. Furthermore, 57 redshift data (42 redshift data from Supernova Cosmology project and 15 redshift data from Cal\'{a}n/ Tolono Supernova survey) are used to estimate the age of the universe and to find the best fit curves for luminosity distance and apparent magnitude.
\end{abstract}

\section{Introduction}
The late time cosmic accelerated expansion of the universe has become one of the most profound researches in modern cosmology. This significant discovery has revealed at the end of the 19th century by modern cosmological observations \cite{Riess09, Perlmutter51, Perlmutter65}. Vigorous investigation on this subject has been going on since then. However, the reason of this acceleration is still unclear. Recently, this result has been considered as one of the greatest significant findings of the 20th century because it contradicts the  general relativity which state that a universe filled with a mixture of ordinary matter and radiation should experience a slowing down expansion.
Two alternative possibilities of accelerated expansion have been profoundly studied: (i)  $75\%$ of the energy density of the universe exists in an unknown form with huge negative pressure known as dark energy   and (ii)  General Relativity (GR) breaks down on cosmological scales and must be replaced with a more complete theory of gravity\cite{Frieman85}.  Subsequently, new theories
and modifications of GR have been suggested to understand this accelerated expansion \cite{Nojiri12, Nojiri15, Nojiri, Harko20, Clifton, Berti, Nojiri:2017ncd}.\\

\noindent
GR has been modified in various different ways and the large number of modified theories are available in the literature \cite{Dvali08, Capozziello83, Nojiri09, Ferraro31, Horava01, Capozziello, Padmanabhan15}. The $f(R)$ theory of modified gravity \cite{Buchdahl} is one of the modified theories that has gained an ample consideration  for its capability to elucidate the accelerated expansion of the universe. In the early 1980s, Starobinsky \cite{Starobinsky99} discussed a most simple $f(R)$ model by taking $f(R)=R+\alpha R^2$, with $\alpha>0$, which is considered as first model representing the inflationary scenario of the universe. Further, the scenario to unify inflation with dark energy in consistent way was proposed in \cite{Nojiri:2003ft}. In this paper, they tried to show that the inflationary epoch can be realized with the term having positive powers of curvature  while the terms with negative powers of curvature indicate to effective dark energy that supports current acceleration.
The $f(R)$ theories of gravity can be formed by modifying the Lagrangian of Einstein Hilbert (EH) action. In EH action, the scalar curvature, $R$, is replaced by an arbitrary function, $f(R)$. Some  $f(R)$ models \cite{Nojiri05, Capozziello35, Nojiri25, Fay, Hu04, Song17, Nojiri38, Nojiri43, Appleby7, Starobinsky57} with the effective cosmological constant phase are very interesting because these may easily reproduce the well-known $\Lambda$CDM cosmology and a subclass of such models which does not violate solar system tests, represents an alternative for standard general
relativity.
A generalization of $f(R)$ gravity suggested in \cite{Bertolami} integrates an explicit coupling between the matter Lagrangian and an arbitrary function of the scalar curvature, which leads to an extra force in the geodesic equation of a perfect fluid. Subsequently, it is shown that this extra force may provide an explanation for the accelerated expansion of the universe \cite{Nojiris, Bertolami1, Bamba}. Apart from this, recently  many authors have investigated the cosmological dynamics in $f(R)$ gravity from various contexts\cite{Capozziello91, Godani01, Bombacigno21, Sbis, Chen, Elizalde26, Elizalde25, Astashenok1, Miranda, Nascimento, Odintsov20, Odintsov1, Nojiri56, Parth, Samanta19, Godani19, Samantaepjc}.  However, unfortunately, $f(R)$ theories may also have some inadequacies. For example, solar system tests have ruled out a good deal of the $f(R)$ models suggested so far \cite{Chiba}. Nevertheless, number of realistic consistent $f(R)$ models which pass solar system tests were proposed in \cite{Cognola:2007zu, Nojiri:2007cq, Nojiri21}.\\

\noindent
A fascinating characteristic of modified theories is the coupling between curvature and matter components. Such type of coupling produces a source term which may yield interesting results and supports to observe the mysteries behind the accelerated expansion of the universe. Motivated by this argument, various modified theories are developed by coupling between matter and curvature components such as $f(R,T)$ gravity\cite{Harko20}, where $T$ symbolizes the trace of energy-momentum tensor. Recently, $f(R, T)$ theory has gained much attention to explain the accelerated expansion of the universe. In
this theory, the matter term $T$ is included in the gravitational action, i.e. the gravitational Lagrangian density is an arbitrary function of both curvature $R$ and matter $T$. The random requirement on $T$ embodies the conceivable contributions from both non-minimal coupling and unambiguous $T$
terms.\\

\noindent
To study the effect of cosmological dynamics in $f(R, T)$ gravity, the numerous functional forms of $f(R, T)$ theories have been taken into consideration in different aspects. The split-up case, $f(R, T)=f_1(R)+f_2(T)$, has received a lot of consideration because one can explore the contributions from
$R$ without specifying $f_2(T)$ and, similarly, one can explore the contributions from $T$ without specifying $f_1(R)$.
Reconstruction of $f(R, T)$ gravity in such separable theories is studied in \cite{Houndjo1}.
A non-equilibrium picture of thermodynamics at the apparent horizon of the (FLRW) universe was discussed in \cite{Sharif01}. Alvarenga et al.\cite{Alvarenga26} studied the evolution of scalar perturbations in $f(R, T)$ gravity with the functional form $f(R, T)=f_1(R)+f_2(T)$. The $f(R,T)$ gravity models satisfying the energy conditions are studied by Alvarenga et al. \cite{Alvarenga}. The inhomogeneity
factors of matter density for self-gravitating
celestial stars are explored using the framework of $f(R,T)$ gravity in \cite{Yousaf2, Bhatti2, Bhatti3, Bhatti4, Yousaf3, Yousaf4}.
Moraes et al. \cite{Moraes5} determined analytical wormhole solutions in $f(R,T)$ theory of gravity. Zubair et al. \cite{Zubair} considered three forms of fluids and analyzed the energy conditions using $f(R,T)$ gravity.
Yousaf et al. \cite{Yousaf1} considered a particular form of matter and obtained wormhole solutions in $f(R,T)$ gravity by taking combinations of shape function.
Bhatti et al. \cite{Bhatti1} obtained the wormhole solutions in modified $f(R,T)$ theory by considering anisotropic, isotropic and barotropic fluid distributions and investigated the energy conditions. Elizalde and Khurshudyan \cite{Elizalde1} considered two types of varying Chaplygin gas and found the dissatisfaction of energy conditions in $f(R,T)$ gravity.  Further, the same authors \cite{Elizalde2} considered the same background for exploration of wormhole solutions by taking different forms of energy density.
Moraes et al. \cite{Moraes1} found the validity of energy conditions for charged wormholes in $f(R, T)$ theory of gravity.
Godani and Samanta \cite{Godani} found the exact wormhole solutions free from exotic matter using $f(R,T)=T+2\alpha \ln(T)$ gravity with $\alpha<0$ and the radius of the throat $>2.7$.
Recently, Godani et al. \cite{Godani1} studied traversable wormhole solutions for $f(R,T)$ gravity model with quadratic  term of $R$ and logarithmic term of trace  $T$ using three novel forms of energy density.
Many authors have been studied deeply this particular form of $f(R, T)=f_1(R)+f_2(T)$, to understand the cosmological dynamics in various contexts\cite{Jamil1, Myrzakulov03, Santos1, Samanta, Shabani, Samanta3, Chandel, Samanta1, Shabani001, Samanta2, Moraes01, Noureen, Farasat, Mirza, Correa1, Moraes9, Zaregonbadi, Rahaman1, Yousaf, Samanta5, Yousaf68, Yousaf07, Elizalde01, Elizalde, Ordines}.\\

\noindent
The motivation of this paper is to study the effect of the cosmological dynamics and estimation of cosmological parameters in viable $f(R, T)$ gravity by considering a functional form of $f(R, T)=f(R, T)=R+\xi T^{1/2}$, where $\xi$ is a constant and $T=\rho-3p$.
The form $f(R,T) = R + \xi T^{1/2}$ satisfies the conservation law and is considered by Velten and  Caram\^{e}s \cite{Velten}  to investigate the cosmological viability of $f(R,T)$ gravity.
They obtained the transition to acceleration for any $\xi>1.2$ and  they found it  closed to the $\Lambda$CDM model for larger values of $\xi$ and smaller values of redshift.
From the above choice of $f(R, T)$, the term $\rho-3p$ must be positive,  in order to have a well defined $f(R, T)$ function. Therefore, the constraint $\rho-3p>0$ is mandatory. The energy conditions are investigated to  assure that our model does not contain any exotic type of matter, for this particular choice of  $f(R, T)$ function. Then the present values of Hubble and deceleration parameters and present age of the universe are estimated. Moreover, the best fit curves are obtained for luminosity distance and apparent magnitude.

\section{$f(R, T)$ Gravity and Field Equations}
The $f(R, T)$ theory of gravity is introduced by Harko et al. \cite{Harko20} in $2011$. They extended standard general theory of relativity by modifying gravitational Lagrangian. The gravitational action in $f(R, T)$ theory is given by
\begin{equation}\label{action2}
S=S_G + S_m=\dfrac{1}{16\pi}\int f(R,T)\sqrt{-g}d^4x +\int \sqrt{-g}\mathcal{L} d^4x,
\end{equation}
where $f(R, T)$ is assumed to be an arbitrary function of $R$ and $T$. Precisely, $R$ is scalar curvature and $T$ is the trace of the energy momentum tensor $T_{\mu\nu}$. The matter Lagrangian density is denoted by $\mathcal{L}$, and the energy momentum tensor is defined as \cite{Landau}:

\begin{equation}\label{}
  T_{\mu\nu}=-\frac{2\delta (\sqrt{-g}\mathcal{L})}{\sqrt{-g}\delta g^{\mu\nu}},
\end{equation}
which yields
\begin{equation}\label{}
  T_{\mu\nu}=g_{\mu\nu}\mathcal{L}-2\frac{\partial \mathcal{L}}{\partial g^{\mu\nu}}.
\end{equation}
The trace $T$ is defined as $T=g^{\mu\nu}T_{\mu\nu}$. Let us define the variation of $T$ with respect to the metric
tensor as
\begin{equation}\label{}
  \frac{\delta (g^{\alpha\beta}T_{\alpha\beta})}{\delta g^{\mu\nu}}=T_{\mu\nu}+\Theta_{\mu\nu},
\end{equation}
where $\Theta_{\mu\nu}=g^{\alpha\beta}\frac{\delta T_{\alpha\beta}}{\delta g^{\mu\nu}}$.
Varying action (\ref{action2}) with respect to the metric tensor $g^{\mu\nu}$ yields
\begin{equation}\label{frt}
f_R(R,T)R_{\mu\nu} -\frac{1}{2}f(R,T)g_{\mu\nu} + (g_{\mu\nu}
\square
-\triangledown_\mu\triangledown_\nu)f_R(R,T)=8\pi T_{\mu\nu} -f_T(R, T)T_{\mu\nu}- f_T(R,T) \Theta_{\mu\nu},
\end{equation}
where $f_R(R,T) \equiv \dfrac{\partial f(R, T)}{\partial R}$ and
$ f_T(R,T) \equiv \dfrac{\partial f(R,T)}{\partial T}.$ Note that, if we take $f(R, T)=R$ and $f(R, T)=f(R)$, then the equations \eqref{frt} becomes Einstein field equations of GR and $f(R)$ gravity, respectively. In the present work, we assume that the stress-energy tensor is defined as
\begin{equation}\label{energy}
  T_{\mu\nu}=(p+\rho)u_\mu u_\nu -p g_{\mu\nu}
\end{equation}
and the matter Lagrangian can be taken as $\mathcal{L}=-p$. The
four velocity $u_{\mu}$ satisfies the conditions $u_{\mu}u^{\mu}=1$ and $u^{\mu}\triangledown_{\nu}u_{\mu}=0$. In this present study, we consider $f(R, T)=f_1(R)+f_2(T)$, where $f_1(R)$ is a function of $R$ and $f_2(T)$ is an arbitrary function of trace of the energy momentum tensor, i. e. $T=\rho-3p$.\\

\noindent
In this paper, the $f(R, T)$ function is defined as
\begin{equation}\label{newfunc}
  f(R, T)=R+\xi T^{1/2},
\end{equation}
where $T=\rho-3p>0$ and $\xi$ is a constant. From the above choice of $f(R, T)$, we came to know that $\rho-3p>0$,  otherwise the function $f(R, T)$ will not be well defined. Therefore, $\rho-3p>0$ is mandatory. \\

\noindent
The space-time of the model is assumed to be  flat Robertson-Walker
metric which is defined as
\begin{equation}\label{metric}
  ds^2=dt^2-a^2(t)(dx^2+dy^2+dz^2).
\end{equation}
 Using equations \eqref{energy}, \eqref{newfunc} and \eqref{metric} in the field equations \eqref{frt}, the explicit form of the field equations are obtained as

\begin{equation}\label{f1}
  3\left(\frac{\dot{a}}{a}\right)^2=8\pi \rho +\frac{\xi(\rho-p)}{\sqrt{\rho-3p}}
\end{equation}
\begin{equation}\label{f2}
  \frac{2\ddot{a}}{a}+\left(\frac{\dot{a}}{a}\right)^2=-8\pi p-\frac{\xi\sqrt{\rho-3p}}{2},
\end{equation}
where $R=6\big[\left(\frac{\dot{a}}{a}\right)^2+\frac{\ddot{a}}{a}\big]$.
 The overhead dot stands for the derivative with respect to time `t'. \\
If we consider $\xi=0$, then the field equations \eqref{f1} and \eqref{f2} reduce to
 \begin{equation}\label{mmf1}
  3\left(\frac{\dot{a}}{a}\right)^2=8\pi\rho,
\end{equation}
\begin{equation}\label{mmf2}
  \frac{2\ddot{a}}{a}+\left(\frac{\dot{a}}{a}\right)^2=-8\pi p.
\end{equation}

\section{Energy Conditions}
The energy conditions (ECs), namely Null Energy Condition (NEC), Weak Energy Condition (WEC), Strong Energy Condition (SEC) and Dominant Energy Condition (DEC) are significant energy conditions. In terms of principal pressures, NEC is defined as $NEC\Leftrightarrow ~~ \forall i, ~\rho+p_{i}\ge 0$; WEC is defined as $WEC\Leftrightarrow \rho\ge 0,$ and $\forall i,  ~~ \rho+p_{i}\ge 0$;  SEC is defined as
$T=\rho+\sum_j{p_j}$ and $SEC\Leftrightarrow \forall j, ~ \rho+p_j\geq 0, ~ \rho+\sum_j{p_j}\geq 0$ and DEC is defined as $DEC\Leftrightarrow \rho\ge 0;$ and $\forall i, ~ p_i\in [-\rho, ~+\rho]$.\\


\noindent
The present work is aimed at the study of FRW model  in the framework of $f(R,T)$ gravity with the non-linear function $f(R,T)=R+\xi\sqrt{T}$, where $\xi$ is an arbitrary constant. First, the field equations are derived in Section-2 which contain three unknown functions of time $t$: (i) scale factor ($a$), (ii) energy density ($\rho$)  and (iii) pressure $(p)$. So, there are two field equations and three unknowns. Thus, one more condition is required for solution of the field equations.  \\

\noindent
We define the scale factor as $a(t)=(t^2+\frac{k}{1-\gamma})^{\frac{1}{3(1-\gamma)}}$,
where $k$ and $\gamma$ are constants.
The aim of defining this scale factor is to solve the field equations \eqref{f1} and \eqref{f2}. Since these field equations are non-linear in $\rho$ and $p$, the exact solution is not possible. Therefore, we have solved these field equations numerically for $\rho$ and $p$ by taking initial value of $\rho$ equal to unity and choosing initial value of $p$ such that $\rho-3p>0$. Then these solutions are utilized in the investigation of energy conditions null, strong and dominant energy conditions.  The scale factor $a(t)=(t^2+\frac{k}{1-\gamma})^{\frac{1}{3(1-\gamma)}}$ consists of two constant $k$ and $\gamma$. The $f(R,T)$ function $f(R,T)=R+\xi\sqrt{T}$ also contains one constant $\xi$. Thus,  there are three constants $\gamma$, $k$ and $\xi$  present in the field equations  \eqref{f1} and \eqref{f2} that can have any real value. We have analyzed the results for energy conditions in three cases  I. $\xi>0$, II. $\xi=0$ and III. $\xi<0$. For each case, there are following 9 possible subcases: 1. $\gamma>0$, $k>0$, 2. $\gamma>0$, $k<0$, 3. $\gamma>0$, $k=0$, 4. $\gamma<0$, $k>0$, 5. $\gamma<0$, $k<0$, 6. $\gamma<0$, $k=0$, 7. $\gamma=0$, $k>0$, 8. $\gamma=0$, $k<0$, 9. $\gamma=0$, $k=0$.
Thus, there are total 27 subcases which are as follows:
\begin{table}[!h]
	\centering
	\begin{tabular}{ccc}
		1.$\xi>0$, $\gamma>0$, $k>0$, & 2. $\xi>0$, $\gamma>0$, $k<0$,& 3. $\xi>0$, $\gamma>0$, $k=0$,\\
		4. $\xi>0$, $\gamma<0$, $k>0$, & 5. $\xi>0$, $\gamma<0$, $k<0$, & 6. $\xi>0$, $\gamma<0$, $k=0$,\\
		7. $\xi>0$, $\gamma=0$, $k>0$, & 8. $\xi>0$, $\gamma=0$, $k<0$, & 9. $\xi>0$, $\gamma=0$, $k=0$,\\
		10.$\xi=0$, $\gamma>0$, $k>0$, & 11. $\xi=0$, $\gamma>0$, $k<0$,& 12. $\xi=0$, $\gamma>0$, $k=0$,\\
		13. $\xi=0$, $\gamma<0$, $k>0$, & 14. $\xi=0$, $\gamma<0$, $k<0$, &15. $\xi=0$, $\gamma<0$, $k=0$,\\
		16. $\xi=0$, $\gamma=0$, $k>0$, & 17. $\xi=0$, $\gamma=0$, $k<0$, & 18. $\xi=0$, $\gamma=0$, $k=0$,\\
		19.$\xi<0$, $\gamma>0$, $k>0$, & 20. $\xi<0$, $\gamma>0$, $k<0$,& 21. $\xi<0$, $\gamma>0$, $k=0$,\\
		22. $\xi<0$, $\gamma<0$, $k>0$, & 23. $\xi<0$, $\gamma<0$, $k<0$, & 24. $\xi<0$, $\gamma<0$, $k=0$,\\
		25. $\xi<0$, $\gamma=0$, $k>0$, & 26. $\xi<0$, $\gamma=0$, $k<0$, & 27. $\xi<0$, $\gamma=0$, $k=0$
	\end{tabular}
\end{table}

\noindent
Since the considered $f(R, T)$ function is well defined only for $T=\rho-3p>0$, we have also examined the nature of $T$ along with the energy conditions. We are discussing them here with respect to positive, zero and negative values of $\xi$ in the following three cases:\\

\noindent
\textbf{Case I. $\xi>0$} \\
In this case, the stress energy tensor $T$ and energy condition terms are either negative or complex for every value of $\gamma$, $k$ and $t$. So $\xi>0$ does not provide the information regarding the presence of normal matter in the universe. \\

\noindent
\textbf{Case II. $\xi=0$} \\
\underline{$\gamma>0$}: If $k>0$, then $T$ and $\rho$ are positive for every value of  $t$ but energy condition terms $\rho+p$, $\rho+3p$ and $\rho-|p|$ are negative. It shows the dissatisfaction of energy conditions. For $k<0$, the ECs are also not satisfied. Further for $k=0$, if we consider $0<\gamma<0.3$, then $T$, $\rho+p$ and  $\rho-|p|$ are positive but $\rho$ and $\rho+3p$ are negative for every $t>0$. Thus  all NEC, DEC and SEC are invalid everywhere. On the other hand, for $\gamma\geq 0.3$ all ECs are also violated. Thus, we have dissatisfaction of ECs for $\xi=0$, $\gamma>0$ and for any value of $k$. \\

\noindent
\underline{$\gamma<0$}: If $k>0$, then $T$ and $\rho$ are positive for  $-1<\gamma<0$ and $t>0$. All $\rho+p$, $\rho+3p$ and $\rho-|p|$ are negative for every value of $t$ and $\gamma$.  Thus all ECs are disobeyed for $k>0$. If $k<0$ or $k=0$, then  all the terms  $T$, $\rho$, $\rho+p$, $\rho+3p$ and $\rho-|p|$  are either negative or complex. Hence, ECs NEC, SEC and DEC are not satisfied for every value of $t$. \\

\noindent
\underline{$\gamma=0$}: If $k>0$, then $T$ and $\rho$ are positive for every value of $t$. $\rho+p$ and $\rho-|p|$ are positive for $t\in(0,0.2]\cup(5.6,\infty)$ and $\rho+3p>0$ for  $t\in(0,0,2]\cup(8.6,\infty)$. Thus, all ECs are valid for $t\in(0,0,2]\cup(8.6,\infty)$. If $k<0$, then $T$ and EC terms are negative. If $k=0$, then $T$ and EC terms are complex. Hence in this subcase, we have obtained the satisfaction of ECs only for $t\in(0,0.2]\cup(8.6,\infty)$ with $\xi=0$, $\gamma=0$, $k>0$.\\

\noindent
\textbf{Case III. $\xi<0$} \\
\underline{$\gamma>0$}: If $k>0$, then ECs are dissatisfied for $\xi<-6$ and $t>0$. Further, for $\xi\geq -6$, $T$ is positive for $t>0$, $\rho>0$ for $t\in(0,0.195)\cup(0.24,\infty)$, $\rho + p\geq0$ for $t\in(0,0.01)\cup(0.37,\infty)$, $\rho +3p\geq0$ for $t\in(0,0.01)\cup(1.1,\infty)$ and $\rho-|p|>0$ for $t\in(0,0.02)\cup(0.37,\infty)$.  Thus, all ECs are obeyed for $t\in(0,0.01]\cup(1.1,\infty)$. If $k<0$, then $T>0$ and  NEC, DEC and SEC are satisfied everywhere for $-54<\xi<-1$, otherwise all ECs are dissatisfied. If $k=0$,  then $T$ and EC terms are complex numbers. Thus, this subcase gives the favorable results for (i)  $t\in(0,0.01]\cup(1.1,\infty)$ with $\xi<0$, $\gamma>0$, $k>0$ and (ii) $t>0$ with $-54<\xi<-1$, $\gamma>0$, $k<0$.  \\

\noindent
\underline{$\gamma<0$}: If $k>0$, then $T>0$ and  all ECs are satisfied everywhere for $-54<\xi<-1.3$, otherwise all ECs are dissatisfied. If $k<0$,  then $T$ and EC terms are complex numbers. If $k=0$, then $T>0$ and  all ECs are valid everywhere for $-8.9<\xi<-1$, otherwise all ECs are invalid. Thus, we have found the obeying nature of the ECs for (i) $t>0$ with $-54<\xi<-1.3$, $\gamma<0$, $k>0$ and (ii) $t>0$ with $-8.9<\xi<-1$, $\gamma<0$, $k=0$\\

\noindent
\underline{$\gamma=0$}: If $k>0$, then  $T>0$ and  NEC, DEC and SEC are satisfied everywhere for $-54<\xi<-1$, otherwise all ECs are dissatisfied.  If $k<0$,  then $T$ and EC terms are complex numbers. If $k=0$, $T>0$ and  NEC, DEC and SEC are satisfied everywhere for $-53<\xi<0$, otherwise all ECs are dissatisfied. Hence ECs are valid for (i) $t>0$ with $-54<\xi<-1$, $\gamma=0$, $k>0$ and (ii) $t>0$ with $-53<\xi<0$, $\gamma=0$, $k=0$.\\

\noindent
Thus, the ranges for the values of constants $\xi$, $\gamma$ and $k$ that provides the validation of the energy conditions and shows the presence of normal matter in the model are  $\xi=0$, $\gamma=0$, $k>0$ (subcase 16); $\xi<0$, $\gamma>0$, $k>0$ (subcase 19); $-54<\xi<-1$, $\gamma>0$, $k<0$ (subcase 20);  $-54<\xi<-1.3$, $\gamma<0$, $k>0$ (subcase 22) $-8.9<\xi<-1$, $\gamma<0$, $k>0$ (subcase 24); $-54<\xi<-1$, $\gamma=0$, $k>0$ (subcase 25) and  $-53<\xi<0$, $\gamma=0$, $k=0$ (subcase 27).  We have plotted $T$, $\rho$, $\rho+p$, $\rho+3p$ and $\rho-|p|$ for subcase 19 with $\xi=-1$, $k=10$ and $\gamma =0.2$ in Fig 1(a)-1(i).\\

\noindent
The results for $\xi=0$ and $\xi<0$ are also summarized in Tables-2 and Table-3 respectively. For  $\xi=0$, the model reduces to GR and provides the satisfaction of energy conditions  for $t\in(0,0.2]\cup(8.6,\infty)$ with $\gamma=0$ and $k>0$. For  $\xi<0$, we obtain the validation of energy conditions (i) for $t\in(0,0.01]\cup(1.1,\infty)$ with $\xi\geq -6$, $\gamma>0$ and $k>0$; (ii) for $t>0$ with $-54<\xi< -1$, $\gamma>0$ and $k<0$; (iii) for $t>0$ with $-54<\xi<-1.3$, $\gamma<0$ and $k>0$; (iv) for $t>0$ with $-8.9<\xi -1$, $\gamma<0$ and $k=0$; (v) for $t>0$ with $-54<\xi< -1$, $\gamma=0$ and $k>0$; (vi) for $t>0$ with $-53<\xi< 0$, $\gamma=0$ and $k=0$. Among these six subcases for $\xi<0$, the energy conditions are fulfilled for every $t>0$ in five subcases. This shows the importance of $f(R,T)$ gravity and confirms the presence of ordinary matter.
 So, there is a large difference in the results for GR and $f(R,T)$ gravity and it may be because of the valid choices of $f(R,T)$ function and scale factor.

\begin{figure}{}
	\centering
	\subfigure[ In this figure, stress energy tensor $T=\rho-3p$ is plotted with respect to $t$. It  is found to be positive for all values of $t$.] {\includegraphics[scale=.67]{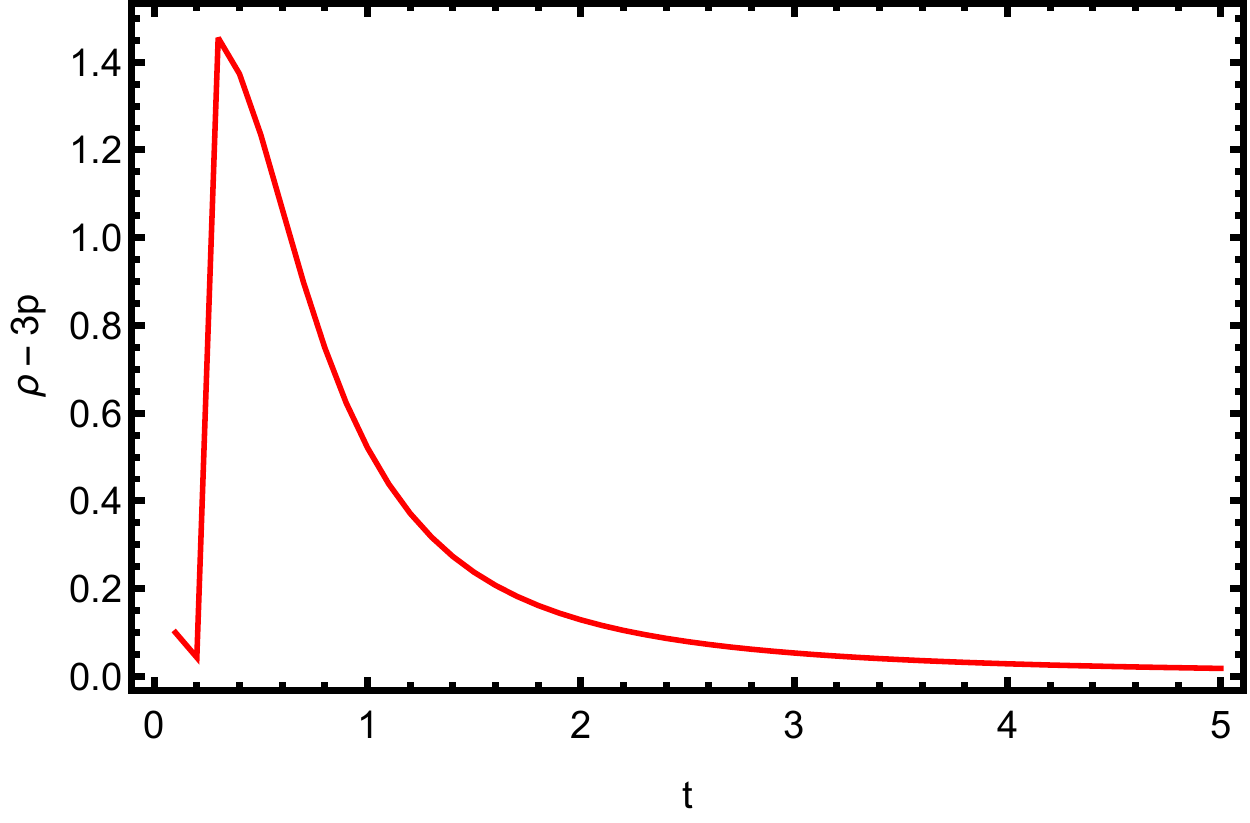}}\hspace{.4cm}
	\subfigure[In this figure, the energy density $\rho$ is plotted for $t\in (0,0.5)$. It is found to be positive for
	$t\in (0,1.95)\cup(0.22,0.5)$.]{\includegraphics[scale=.67]{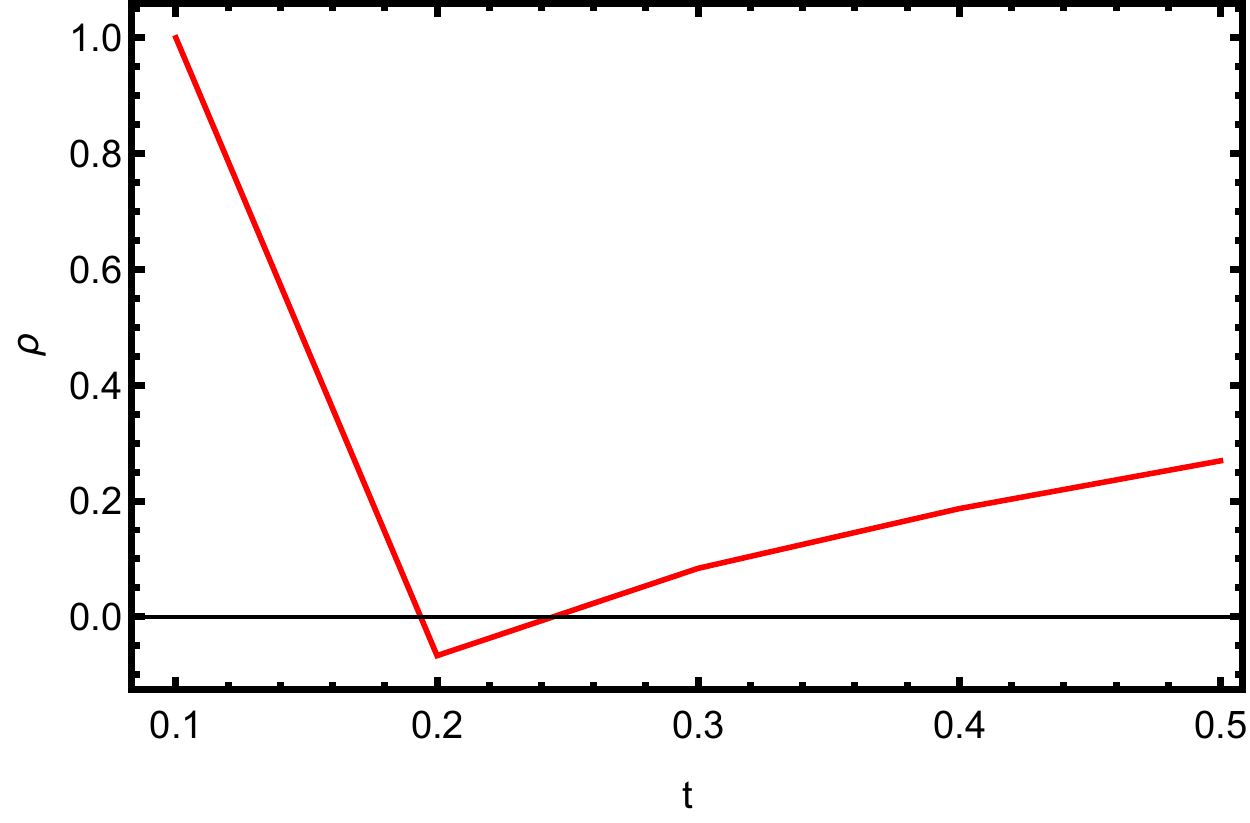}}\hspace{.4cm}
	\subfigure[In this figure, the energy density $\rho$ is plotted for $t\geq0.5$. It is found to be a positive and decreasing function for every $t\geq0.5.$]{\includegraphics[scale=.67]{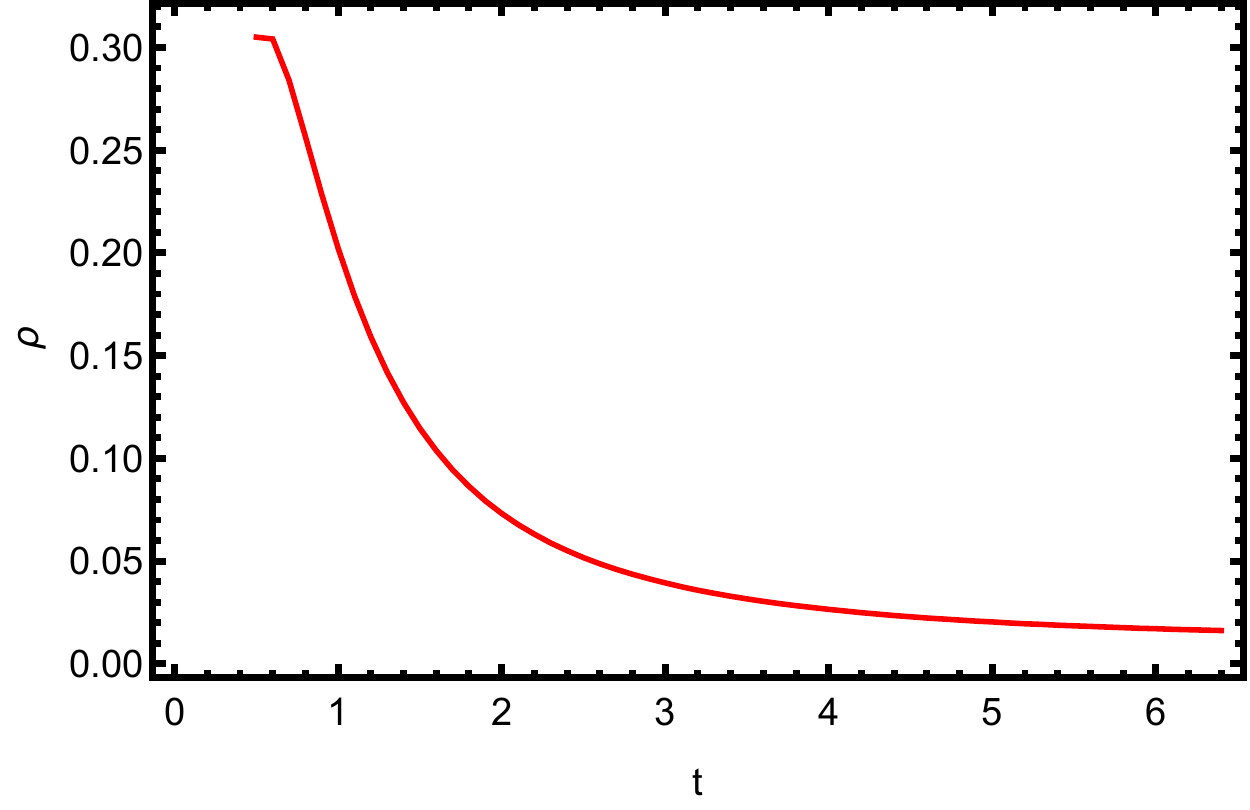}}\hspace{.4cm}
	\subfigure[In this figure, the NEC term $\rho+p$ is plotted for $t\in(0,0.4)$. It is found to be positive for $t\in(0,0.02)\cup(0.37,0.4)$.] {\includegraphics[scale=.67]{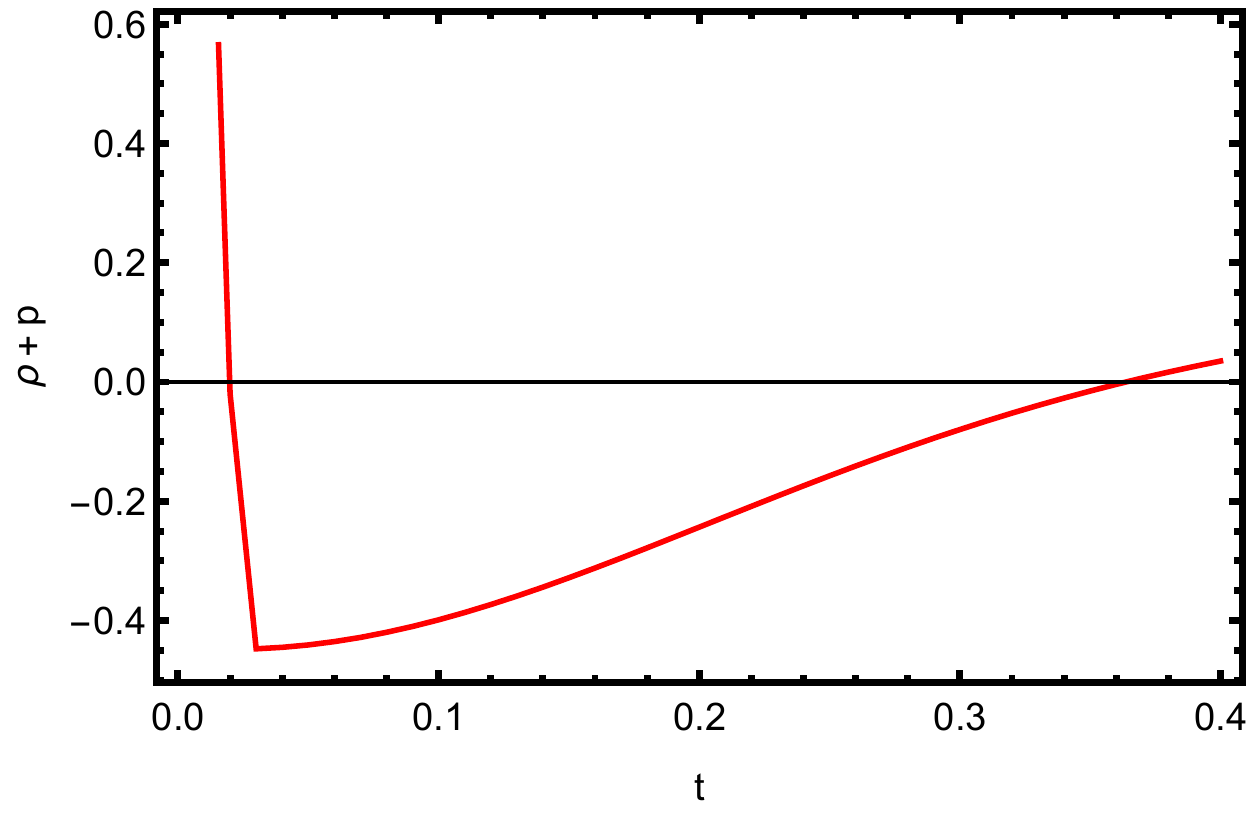}}\hspace{.4cm}
	\subfigure[In this figure, the NEC term $\rho+p$ is plotted for $t\geq0.4$. It is found to be a positive and decreasing function for every $t\geq0.4$.] {\includegraphics[scale=.67]{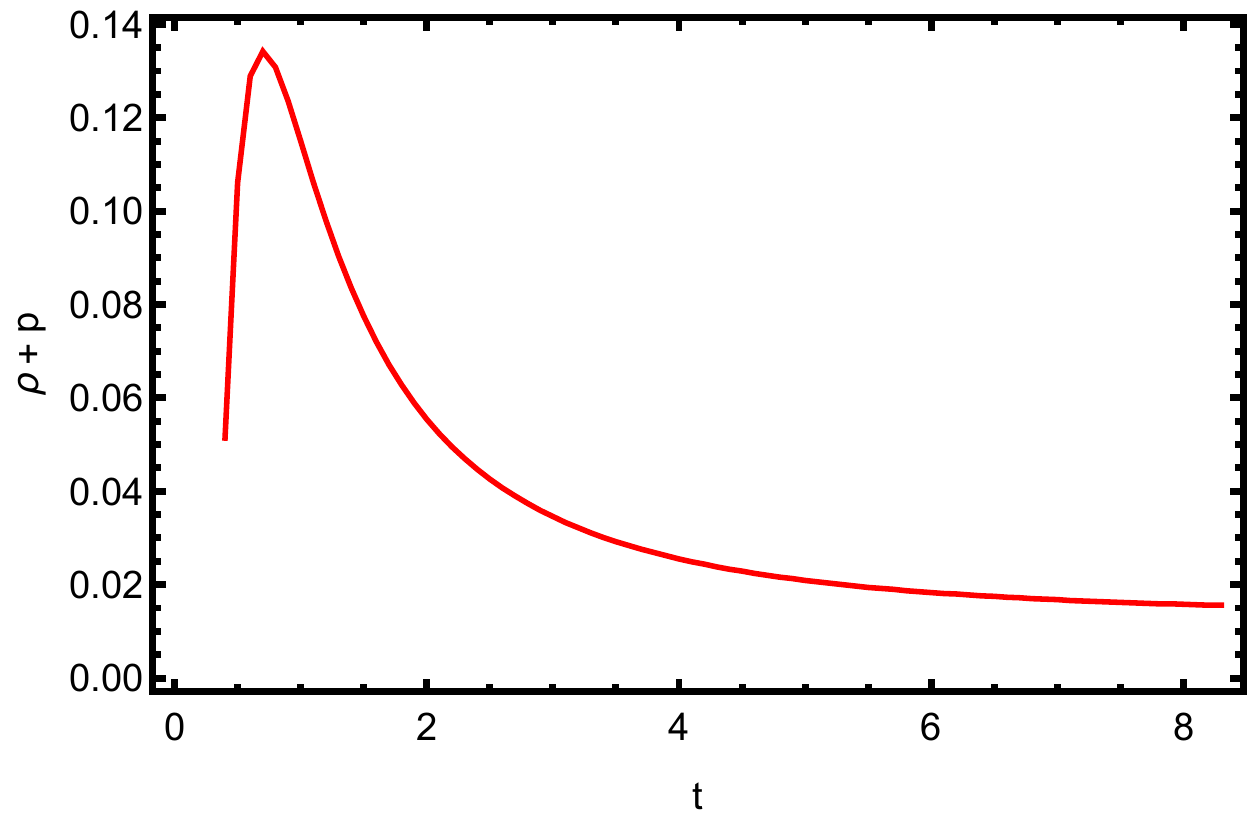}}\hspace{.4cm}
	\subfigure[In this figure, the SEC term $\rho+3p$ is plotted for $t\in(0.1.3)$. It is found to be a positive function for every $t\in(0,0.1)\cup(1.1,1.3)$.] {\includegraphics[scale=.67]{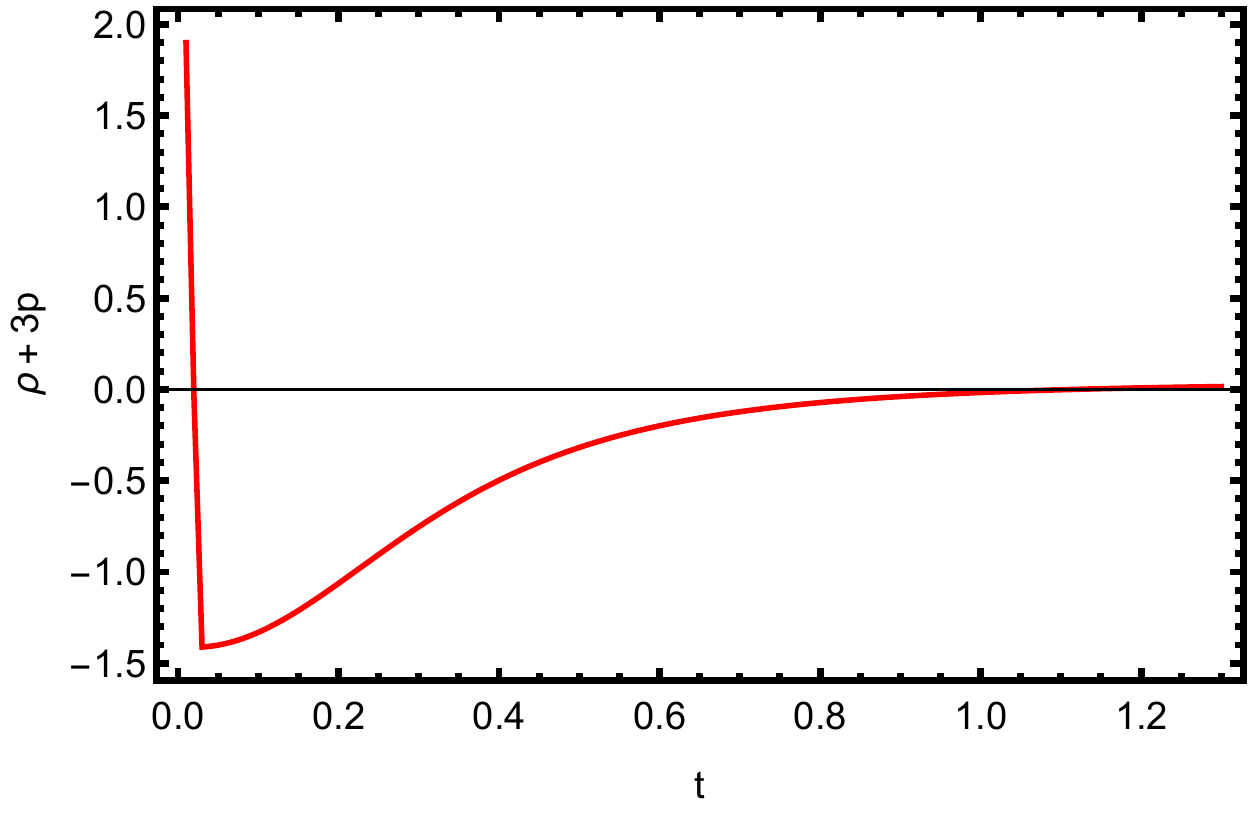}}\hspace{.4cm}
\end{figure}

\begin{figure}{}
	\centering
	\subfigure[In this figure, the SEC term $\rho+3p$ is plotted for $t\geq1.3$. It is found to be a positive function for every $t\geq1.3$.] {\includegraphics[scale=.67]{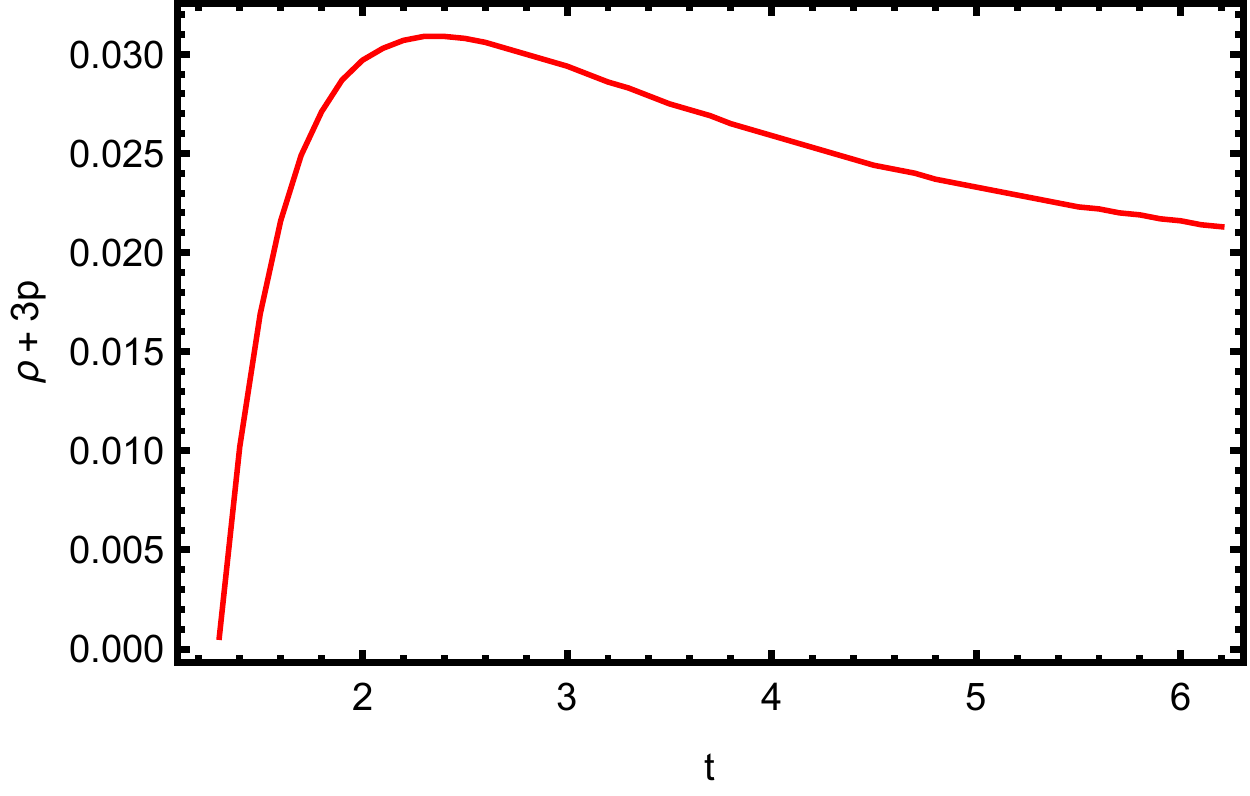}}\hspace{.4cm}
	\subfigure[In this figure, the DEC term $\rho-|p|$ is plotted for $t\in(0,0.4)$. It is found to be a positive function for every $t\in(0,0.02)\cup(0.37,0.4)$.] {\includegraphics[scale=.67]{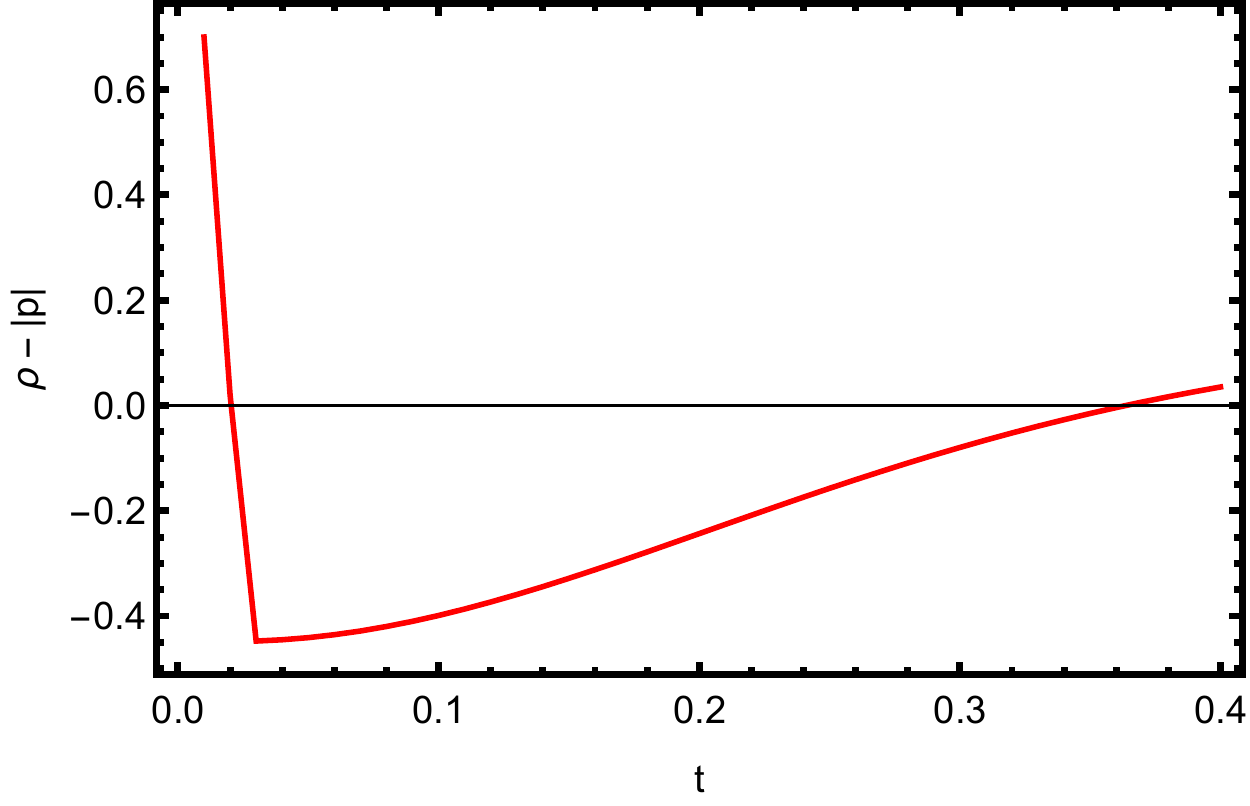}}\hspace{.4cm}
	\subfigure[In this figure, the DEC term $\rho-|p|$ is plotted for $t\geq 0.4$. It is found to be a positive function for every $t\geq 0.4$.] {\includegraphics[scale=.67]{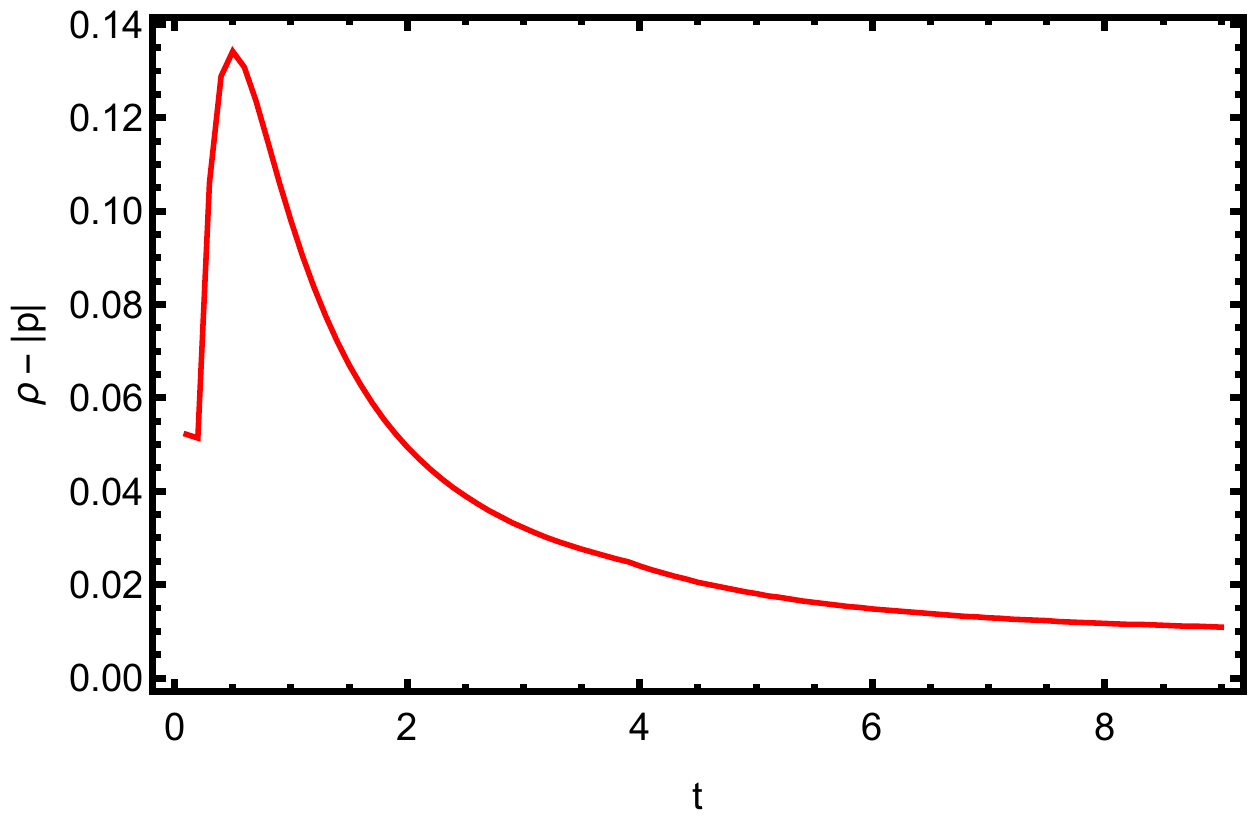}}\hspace{.4cm}
	\caption{Plots for $\rho-3p$, $\rho$, $\rho+p$, $\rho+3p$ and $\rho-|p|$  with $\xi=-1$, $k=10$ and $\gamma =0.2$.  }
\end{figure}

\section{Estimation of Cosmological Parameters}
In this section we would like to estimate some cosmological parameters and analyze our estimated value with observational data.

\subsection{Hubble and Deceleration Parameters}
%

\noindent
In Section 3, the scalar factor $a(t)$ is defined to solve the field equations and analyze the energy conditions. In this section, the same scale factor is used to obtain the expressions for Hubble parameter and deceleration parameter in terms of cosmic time as well as redshift. Further, the present values of these parameters are also estimated.

\noindent
The Hubble parameter ($H$) gives the expansion rate of the universe. It is defined as $H=\dfrac{\dot{a}}{a}$ and can be positive or negative. Its positive value depicts the expanding universe while its negative value represents the contracting universe.
In terms of the cosmic time, it is calculated as
\begin{equation}
H(t)=\frac{2t}{3(t^2+\frac{k}{1-\gamma})}.
\end{equation}
In terms of red shift, it comes out to be
\begin{equation}\label{h}
H(z)=\frac{2}{3s(1-\gamma)}\Big[s-\frac{k}{1-\gamma}(1+z)^{3(1-\gamma)}\Big]^{1/2}(1+z)^{\frac{3(1-\gamma)}{2}},
\end{equation}
where $s=t_0^2+\frac{k}{1-\gamma}$ and $t_0$ is the present age of the universe.\\

\noindent
The deceleration parameter represents the decelerating or accelerating nature of the universe. It is defined as $q=-\dfrac{\ddot{a}}{\dot{a}H}$. Its value lies between -1 and 1. If it lies between 0 and 1, then it means that the universe is under decelerating phase and there is domination of matter over dark energy. If its value lies between -1 and 0, then it means that the universe is under accelerating phase and there is domination of dark energy. Recent observations have supported the accelerating nature of our universe.
In terms of the cosmic time, it is calculated as
\begin{equation}
q(t)=\frac{3k}{2t^2}+\frac{3\gamma}{2}-\frac{1}{2}.
\end{equation}
In terms of red shift, it comes out to be
\begin{equation}
q(z)=\frac{3k}{2}\Big[\frac{s}{(1+z)^{3(1-\gamma)}}-\frac{k}{1-\gamma}\Big]^{-1}+\frac{3\gamma}{2}-\frac{1}{2},
\end{equation}
where $s$ and $t_0$ are same as above.

\begin{figure}[h!]\label{hh}
	\begin{center}
		\includegraphics{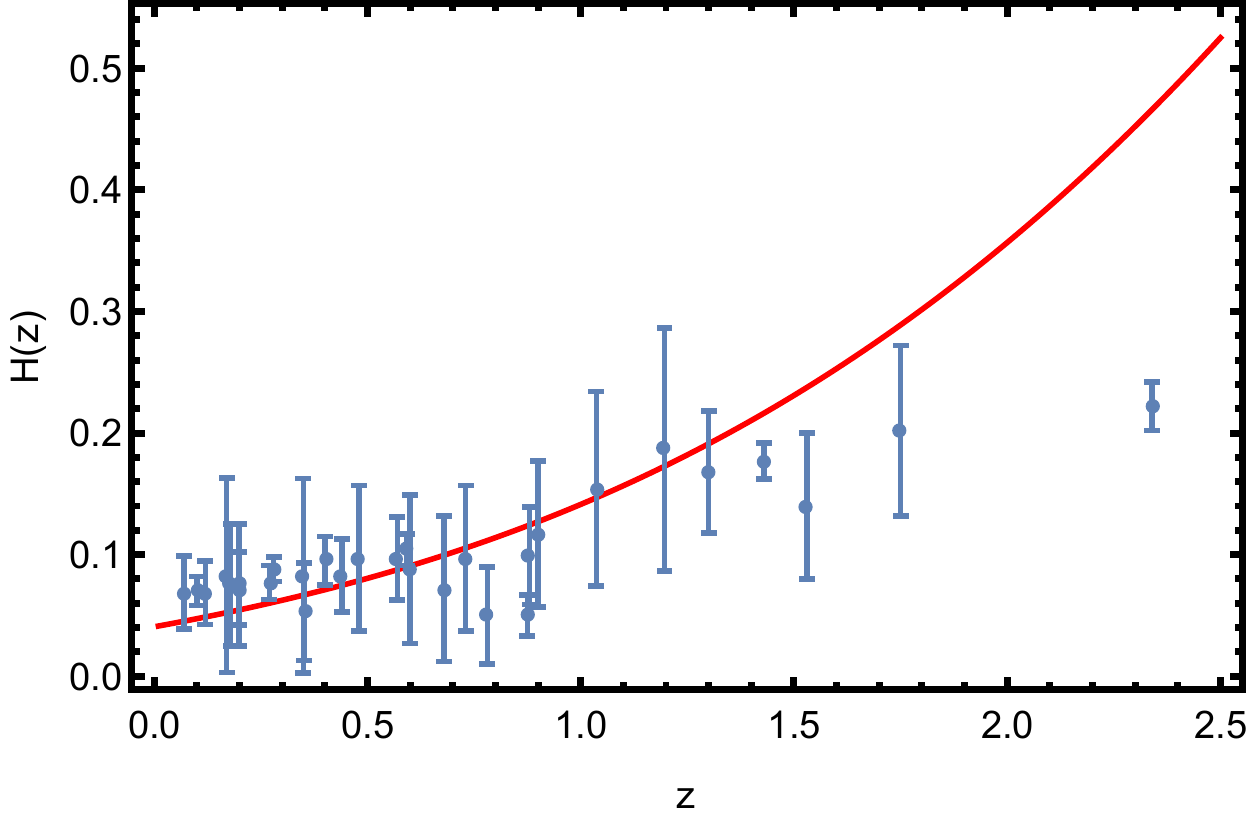}
		\caption{In this figure, error bar graph is drawn for Hubble parameter $H(z)$ verses redshift $z$ with $k=-45.73$ and $\gamma=0.06542$. The red curve is drawn for theoretical values of $H(z)$. The blue dots shows the observational values of Hubble parameter and bar on these dots represents the standard error.}
	\end{center}
\end{figure}
\begin{figure}[h!]\label{qq}
	\begin{center}
		\includegraphics{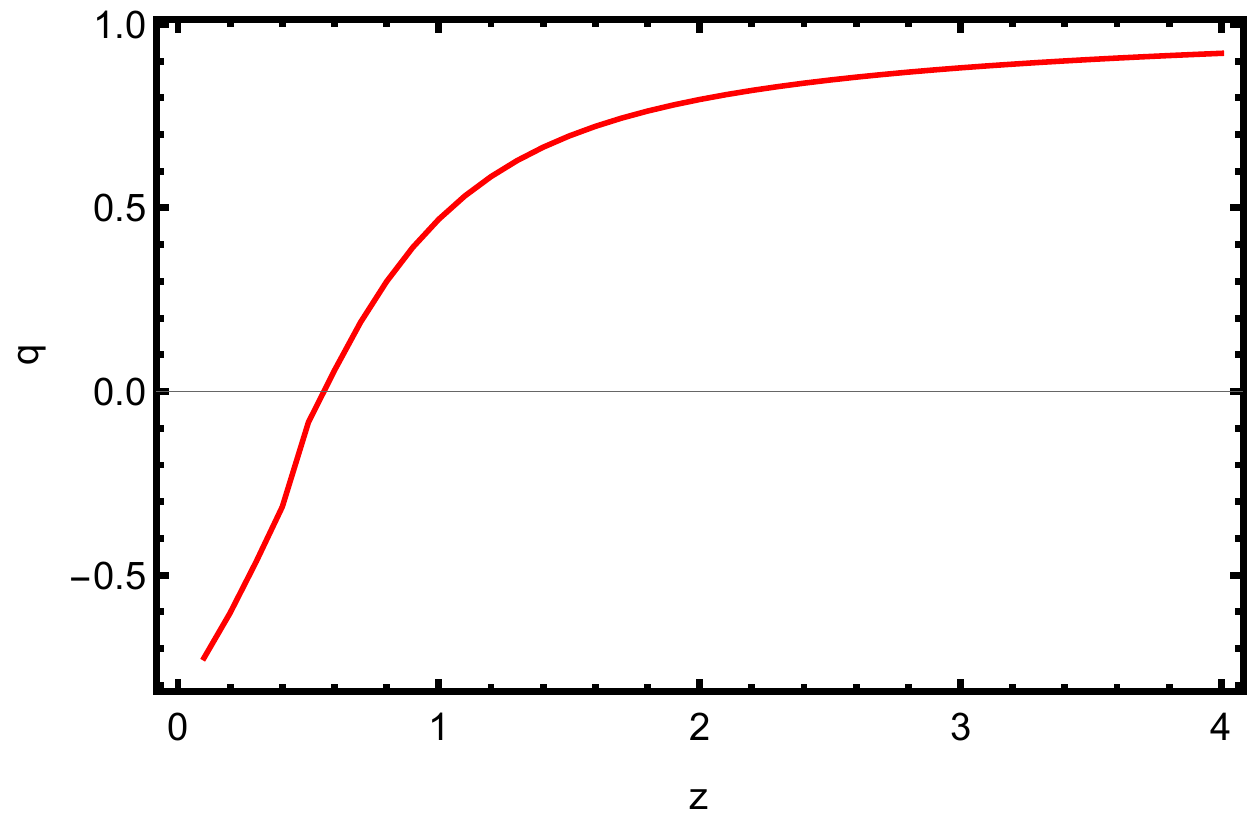}
		\caption{In this figure, the  deceleration parameter $q(z)$ versus redshift $z$ is drawn with $k=-45.73$ and $\gamma=0.06542$. It shows the evolution of the universe from decelerating phase to accelerating phase.}	\end{center}
\end{figure}

\noindent
Both $H(z)$ and $q(z)$ contains two constant $k$ and $\gamma$. In order to obtain an expanding and accelerating universe, the value of these constants are determined as $k=-45.73$ and $\gamma=0.06542$. For these values of $k$ and $\gamma$, the plots for $H(z)$ and $q(z)$ with respect to redshift $z$ are shown in Figs. (1) \& (2). In Fig. (1), the red curve is drawn for theoretical values of $H(z)$ and blue dots  represent the observational values of $H(z)$ mentioned at the end of this article.   At $z=0$, the present value of Hubble parameter is estimated as 0.071413 GYrs$^{-1}$ which is very closed to its present observational value \cite{wmap}. In Fig. (2), the deceleration parameter $q(z)$ is plotted with respect to redshift. It is found to be an increasing function with respect to redshift. Its value is observed to lie between -1 and 1. At $z=0$, its present value is obtained to be equal to -0.725  which is closed to the current experimental value \cite{Rapetti}. Hence, the model represents an expanding  universe with accelerated rate of expansion. Consequently, it shows the significance of the considered scale factor.

\subsection{Luminosity Distance, Apparent Magnitude \& Age of the Universe}	

\noindent
According to the observations of type Ia Supernova \cite{Riess09, Perlmutter65}, the universe is expanding in an accelerating way. Because of such expansion, the stellar objects are redshifted, when they emit light. The luminosity distance is defined by the luminosity of an stellar object. It is the distance measure which is obtained from the  Supernova using the distance modulus. The relation between redshift and luminosity distance is one of the important tools of cosmology to explore the evolution of the universe. Further, the apparent magnitude of a source is associated with the luminosity distance. Another cosmological consequence is the age of the universe which means that how old is the universe? Astronomers  can  estimate it in two ways: 1. by determining the age of the oldest stellar objects, 2. by determining the rate of expansion of the universe. Many cosmologists have calculated the present age of the universe \cite{Jimenez, Richer, Hansen}. According to the WMAP3 data, it is equal to $t_0=13.73^{+.13}_{-.17}$GYrs. In the present section, the expressions for luminosity distance,  apparent magnitude and  age of the universe are determined and the theoretical results are  compared with the corresponding observational results.\\

\noindent
The luminosity distance is defined as
\begin{eqnarray}\label{d1}
D_L&=&a_0c(1+z)\int_{t}^{t_0}\dfrac{dt}{a(t)}\nonumber\\
&=&c(1+z)\int_{0}^{z}\dfrac{dz}{H(z)},
\end{eqnarray}
where $c$ and $a_0$ are the speed of light and the present value of the scale factor respectively.
From Eqns. (\ref{h}) and \eqref{d1},
\begin{eqnarray}\label{dl}
D_L&=c(1+z)\int_{0}^{z}\Big[\frac{2}{3s(1-\gamma)}\big(s-\frac{k}{1-\gamma}(1+z)^{3(1-\gamma)}\big)^{1/2}(1+z)^{\frac{3(1-\gamma)}{2}}\Big]^{-1}dz.
\end{eqnarray}
Let $m$ and $M$ stand for the apparent and absolute magnitudes respectively. Then the relation between these two magnitudes is given as
\begin{equation}\label{mm}
m-M=5log_{10}\Big(\dfrac{D_L}{Mpc}\Big)+25.
\end{equation}
To find the absolute magnitude, it is considered at very low redshift. For lower redshift,
\begin{equation}\label{d}
D_L=\dfrac{cz}{H_0}.
\end{equation}
Using Eq.(\ref{d}) and substituting $z=0.026$ and $m=16.08$ in (\ref{mm}),
\begin{equation}\label{M}
M=5log_{10}\left(\dfrac{H_0}{0.026c}\right)-8.92.
\end{equation}
From Eqs. (\ref{mm}) \& (\ref{M}),
\begin{eqnarray}\label{m1}
m&=&16.08+5log_{10}\left(\dfrac{D_LH_0}{0.026c}\right).
\end{eqnarray}

\noindent
Now, using Eqns. \eqref{h}, \eqref{d1} and \eqref{m1}
\begin{eqnarray*}
m &= &16.08+5log_{10}\left(\dfrac{(1+z)H_0}{0.026}\int_{0}^{z}\dfrac{dz}{H(z)}\right)\\
&= &16.08+5 log_{10}\left(\dfrac{(1+z)H_0}{0.026}\int_{0}^{z}\Big[\frac{2}{3s(1-\gamma)}
\big(s-\frac{k}{1-\gamma}(1+z)^{3(1-\gamma)}\big)^{1/2}(1+z)^{\frac{3(1-\gamma)}{2}}\Big]^{-1}dz\right).
\end{eqnarray*}
\begin{eqnarray}\label{m}
\end{eqnarray}

\noindent
The age of the universe is given as
\begin{equation}\label{t}
t_0=\int_0^{t_0}dt=\int_{0}^{\infty}\dfrac{dz}{H(z)(1+z)}.
\end{equation}
Using Eq. (\ref{h}),
\begin{equation}\label{t}
t_0= \int_0^{t_0}dt=\int_{0}^{\infty}\dfrac{dz}{
	\frac{2}{3s(1-\gamma)}\Big[s-\frac{k}{1-\gamma}(1+z)^{3(1-\gamma)}\Big]^{1/2}(1+z)^{\frac{5-3\gamma}{2}}}.
\end{equation}

\begin{figure}[h!]\label{distance}
	\begin{center}
		\includegraphics{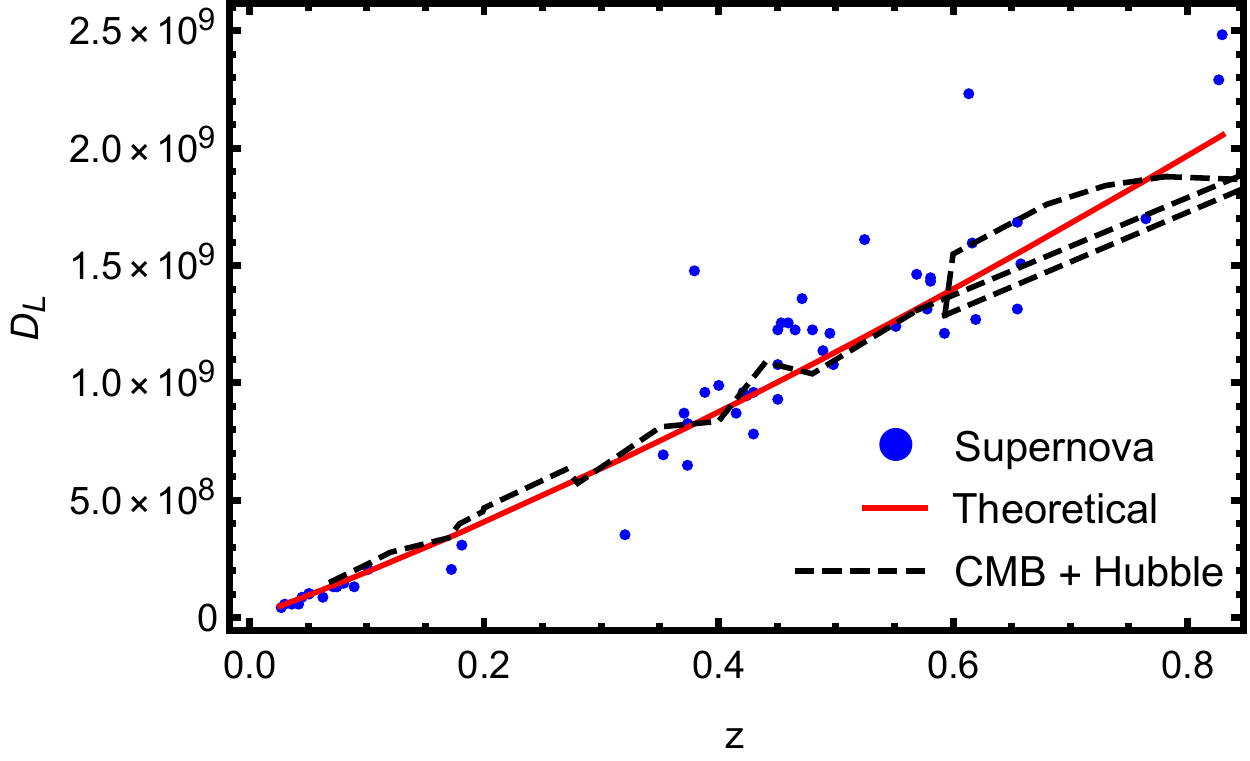}
		\caption{The  luminosity distance $D_L$ versus redshift $z$ }
	\end{center}
\end{figure}

\begin{figure}[h!]\label{ap}
	\begin{center}
		\includegraphics{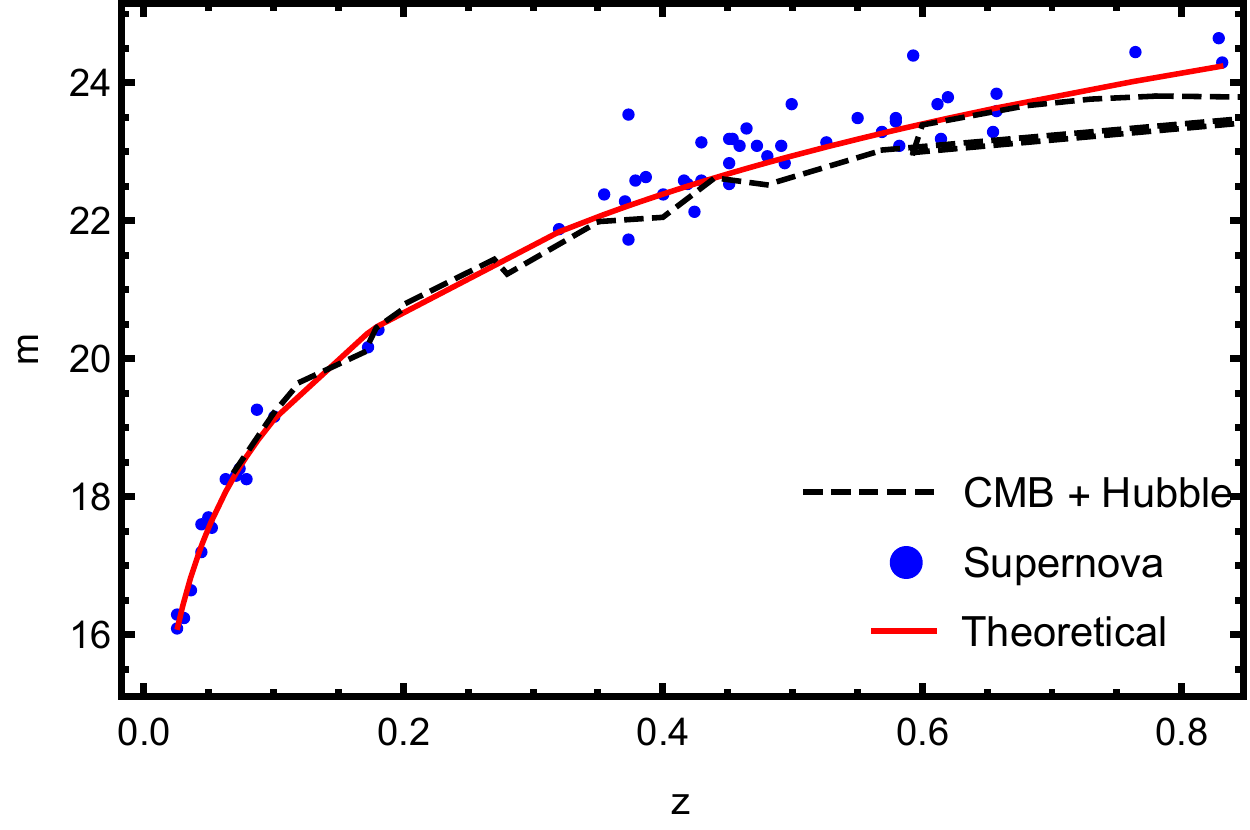}
		\caption{The apparent magnitude versus redshift $z$}
	\end{center}
\end{figure}

\begin{figure}[h!]\label{ag}
	\begin{center}
		\includegraphics{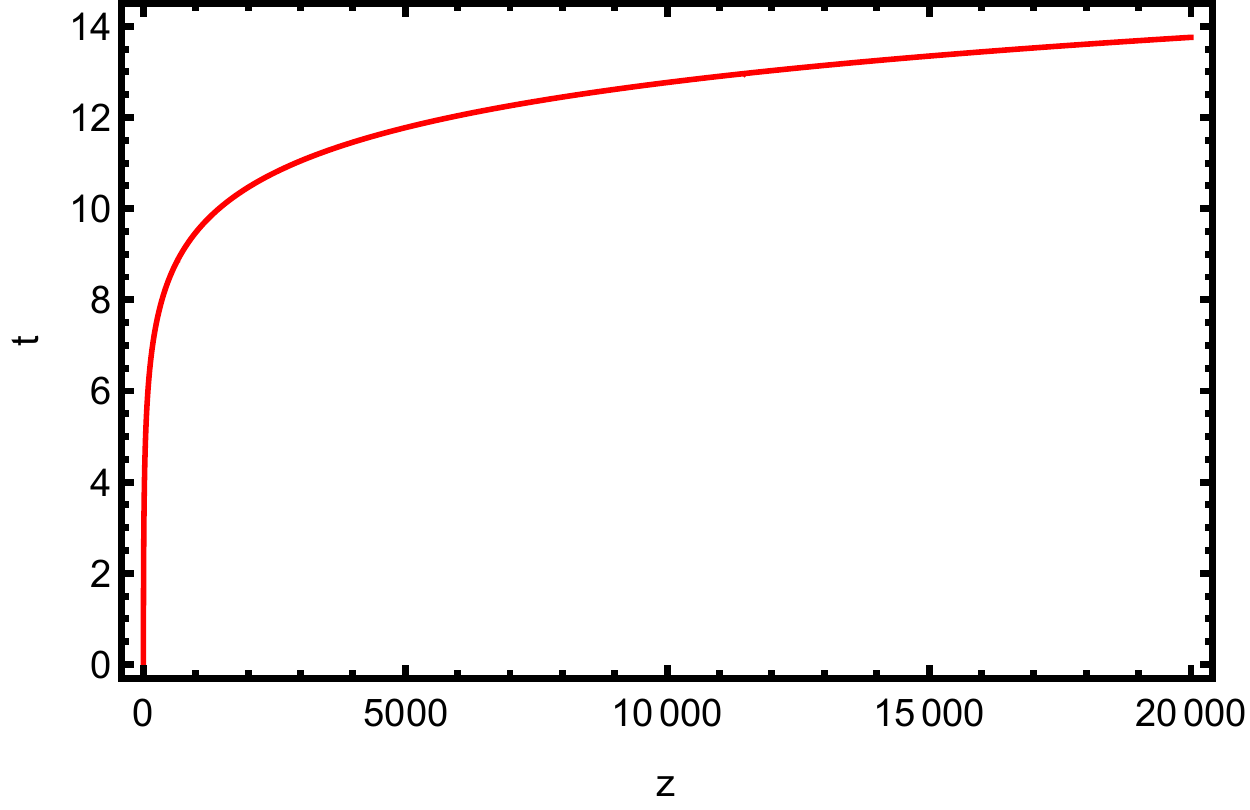}
		\caption{$t$ (in GYrs)  versus redshift $z$ }
	\end{center}
\end{figure}

\noindent
The integrations of \eqref{dl}, \eqref{m} and \eqref{t} give the  luminosity distance, apparent magnitude  and age of the universe respectively.
  Fig. (3) is drawn for luminosity distance $D_L$ with respect to redshift $z$. In this figure, the red solid curve is drawn corresponding to the theoretical values of luminosity distance, blue dots represent its observational values corresponding to Supernove data and black dashes represent its values corresponding to CMB and Hubble data.  In Fig. (4), apparent magnitude $m$ is drawn with respect to redshift $z$. The solid red curve is drawn for its theoretical values, blue dots are marked with respect to Supernove data and black dashes represent its values corresponding to CMB and Hubble data. In both figures, (Figs. (3)\& (4)), we have obtained a best fitting of the curves that shows a consistency between observational data and theoretical results. The theoretical results are in good agreement with the results of Hubble and CMB data in comparison of Supernova data. Further in Fig. (5), the curve is drawn for the cosmic time $t$ with respect to redshift $z$. It is found to be an increasing function and approaches to 13.757. This gives the age of the universe  $t_0$ as 13.757 GYrs. According to WMAP data, the age of our universe is 13.73 GYrs  approximately \cite{wmap}. So, our estimated age is very closed to observational result.

  \section{Causality and Sound Speed}

  Three types of particles are available in the universe: sub-luminal, luminal
and super-luminal. The sub-luminal particles move very slow in comparison of the speed of light,  for example electrons and neutrons. The luminal
particles move with exactly the same speed
as the speed of light, for example photon and graviton. However, the particles
moving  faster than the speed of light are called super-luminal particles or tachyons. There are
two possibilities for the existence of super-luminal particles: either they do not exist or if they
do, then they do not interact with an ordinary matter. If the speed of sound is less than the
local light speed, $c_s\le 1$, then only we can say about the non-violation of causality. The positive square
sound speed ($c_s^2>0$) is necessary for the classical stability of the universe. The speed of sound is
defined as $\frac{dp}{d\rho}=c_s^2$ \cite{Ellis2007}.
We obtain the speed of sound as
\begin{equation}\label{27}
  \frac{dp}{d\rho}=c_s^2= \frac{1}{-32\pi+3\xi}\Big(\frac{AB}{C}+\xi\Big),
\end{equation}
where $A=8\Big(\frac{\dddot{a}}{a}-\frac{\dot{a}^3}{a^3}\Big)\sqrt{\rho-3p}$,
$B=-\xi^2(\rho+3p)-\{16\pi(\rho-3p)^{3/2}+\xi(\rho-5p)\}(-32\pi+3\xi)$,
$C=B\xi(\rho+3p)-D(-32\pi+3\xi)$ and $D=12\Big(\frac{\dot{a}\ddot{a}}{a^2}-\frac{\dot{a}^3}{a^3}\Big)(\rho-3p)^{3/2}$.

In Equation \eqref{27}, $  \frac{dp}{d\rho}$ depends on variable $t$ and constants $\xi$, $k$ and $\gamma$. These constants can be positive or negative. For $\xi>0$, there are four possible cases: $1.$ $k>0, \gamma>0$, $2.$ $k>0, \gamma<0$, $3.$ $k<0, \gamma>0$, $4.$ $k<0, \gamma<0$. Similarly, there are four possible cases for $\xi<0$. Thus, total eight cases are:
$1.$ $\xi>0$, $k>0, \gamma>0$, $2.$ $\xi>0$, $k>0, \gamma<0$, $3.$ $\xi>0$, $k<0, \gamma>0$, $4.$ $\xi>0$, $k<0, \gamma<0$, $5.$ $\xi<0$, $k>0, \gamma>0$, $6.$ $\xi<0$, $k>0, \gamma<0$, $7.$ $\xi<0$, $k<0, \gamma>0$, $8.$ $\xi<0$, $k<0, \gamma<0$.  Further, we have estimated the value of $\frac{dp}{d\rho}$ for  eight  these cases in Table-1.

\begin{table}[!h]
	\centering
	\caption{Results for $  \frac{dp}{d\rho}$}
	\begin{tabular}{|c|c|c|c|}
			\hline
\textbf{S.No.}& $\xi$, $k$ and $\gamma$ & $\xi$ &$E\equiv \frac{dp}{d\rho}$\\\hline
1 & $\xi>0$, $k>0$, $\gamma>0$, $\gamma\neq 1$ & $\xi\in(0,4]$&$E\in(-1,0)$, for all $t$\\
   & & $\xi\in[4,\infty)$&$E>1$, for all $t$\\\hline
   2 & $\xi>0$, $k>0$, $\gamma<0$ & $\xi\in(0,24)$&$E<-1$, for all $t$\\
   & &$\xi\in[24,\infty)$ &$E>1$, for all $t$\\\hline
   3 & $\xi>0$, $k<0$, $\gamma>0$, $\gamma\neq 1$ & $\xi\in(0,3)$&$E\in(-1,0)$, for $t\in(4,10)\cup(16,\infty)$\\
   & & &$E\in(0,1)$, for $t\in(0,4]\cup[10,16]$\\
   & &$\xi\in[3,18)$ &$E<-1$, for all $t$\\
   & &$\xi\in[18,\infty)$ &$E>1$, for all $t$\\
   \hline
   4 & $\xi>0$, $k<0$, $\gamma<0$ & for all $\xi$&$E\in(0,1)$, for all $t$\\
    \hline
    5 & $\xi<0$, $k>0$, $\gamma>0$, $\gamma\neq 1$ & $\xi\in(-8,-1]$&$E\in(-1,0)$, for $t\in(0,0.9]$\\
   & & &$E\in(0,1)$, for $t\in(0.9,\infty)$\\
   & &$\xi\in(-1,0)\cup(-\infty,-8]$ &$E\in(0,1)$, for all $t>0$\\
   \hline
    6 & $\xi<0$, $k>0$, $\gamma<0$ & $\xi\in(-1,0)$&$E\in(-1,0)$, for $t\in(0,0.3)\cup(0.8,\infty)$\\
   & & &$E\in(0,1)$, for $t\in[0.3,0.8]$\\
   &  & $\xi\in(-8,-1]$&$E\in(-1,0)$, for $t\in(0,0.3)$\\
   & & &$E\in(0,1)$, for $t\in[0.3,\infty)$\\
   &  & $\xi\in(-\infty.-8]$&$E\in(0,1)$, for $t\in(0,\infty)$\\
    \hline
     7 & $\xi<0$, $k<0$, $\gamma>0$, $\gamma\neq 1$ & $\xi\in(-1,0)$&$E\in(0,1)$, for $t\in(0,0.3)\cup(0.8,\infty)$\\
    & & &$E\in(-1,0)$, for $t\in[0.3,0.8]$\\
    &  & $\xi\in(-\infty, -1]$&$E\in(0,1)$, for $t\in(0,\infty)$\\
    \hline
    8 & $\xi<0$, $k<0$, $\gamma<0$ & $\xi\in(-1,0)$&$E\in(-2,-1)$, for $t\in(0,0.8)$\\
    & & &$E\in(-1,0)$, for $t\in[0.8,\infty)$\\
    &  & $\xi\in(-\infty,-1]$&$E\in(-1,0)$, for $t\in(0,0.2)$\\
    & & &$E\in(0,1)$, for $t\in[0.2,\infty)$\\
    \hline
\end{tabular}
\end{table}

\noindent
From Table-1, it is observed that the sound speed $\frac{dp}{d\rho}$ is greater than one for (1) $\xi\ge 4$, $k>0$ and $\gamma>0$ throughout the
evolution, which indicates the presence of abnormal matter in the universe. Furthermore, the sound speed $\frac{dp}{d\rho}<0$ for $0<\xi< 4$, $k>0$ and $\gamma>0$, which is not acceptable. So, causality could be violated for $\xi\ge 4$, $k>0$ and $\gamma>0$. For (2) $\xi>0, k>0$ and $\gamma<0$ and (3)
$\xi>0, k<0$ and $\gamma>0$, the sound speed is greater than one, if  $\xi\ge 24$ and $\xi\ge 18$, respectively, throughout the evolution, which indicates the availability of abnormal matter in the universe. However, $\frac{dp}{d\rho}<-1$  indicates the non-availability or presence of ordinary matter in the universe. Similarly, for other range of parameters (4) $\xi>0, k<0, \gamma<0$;
(5) $\xi<0, k>0, \gamma>0$; (6) $\xi<0, k>0, \gamma<0$; (7) $\xi<0, k<0, \gamma>0$ and (8) $\xi<0, k<0, \gamma<0$; the sound speed is less than one throughout the
evolution of the universe for different range of $xi$.

\section{Conclusion}
In this paper, we have investigated FRW model to get an accelerating and expanding universe filled with the non-exotic matter that possesses positive energy density and fulfills the energy conditions. To obtain such universe, we have taken into account the framework of $f(R,T)$ theory of gravity and introduced $f(R, T)=R+\xi T^{1/2}$, where $\xi$ is a constant. Further, we have derived the field equations and defined the scale factor in the form of $a(t)=(t^2+\frac{k}{1-\gamma})^{\frac{1}{3(1-\gamma)}}$, where $k$ and $\gamma$ are arbitrary constants. Since the field equations are non-linear in $\rho$ and $p$, the exact solution is not possible and we have found their numerical solution. Using these solutions, we have analyzed energy conditions for all possible combinations of the values of constants $\xi$, $\gamma$ and $k$. This analysis is done mainly in three cases: $\xi>0$, $\xi=0$ and $\xi<0$. For  $\xi>0$, the energy conditions are violated for every $t>0$. For  $\xi=0$, the model reduces to GR and provides the satisfaction of energy conditions  for $t\in(0,0.2]\cup(8.6,\infty)$ with $\gamma=0$ and $k>0$. Finally for  $\xi<0$, we obtain the validation of energy conditions (i) for $t\in(0,0.01]\cup(1.1,\infty)$ with $\xi\geq -6$, $\gamma>0$ and $k>0$; (ii) for $t>0$ with $-54<\xi< -1$, $\gamma>0$ and $k<0$; (iii) for $t>0$ with $-54<\xi<-1.3$, $\gamma<0$ and $k>0$; (iv) for $t>0$ with $-8.9<\xi -1$, $\gamma<0$ and $k=0$; (v) for $t>0$ with $-54<\xi< -1$, $\gamma=0$ and $k>0$; (vi) for $t>0$ with $-53<\xi< 0$, $\gamma=0$ and $k=0$. Among these six subcases for $\xi<0$, the energy conditions are fulfilled for all $t>0$ in five subcases. This shows the importance of $f(R,T)$ gravity and assures the presence of non-exotic matter. Thereafter, we have used 31 observational data to fit the curve for Hubble parameter and determined its present value as 0.071413 GYrs$^{-1}$. Then deceleration parameter is obtained showing the evolution of the universe from decelerating phase to accelerating phase and at $z=0$, it is calculated as $-0.725$. Further, experimental  data for luminosity distance and apparent magnitude are used to best fit the corresponding curves obtained from theoretical values. Furthermore, a curve between $t$ and $z$ is drawn which approaches to 13.757 as redshift $z$ approaches to infinity. It gives the present age of the universe as 13.757 GYrs. Finally, we conclude that all the results are well consistent with the corresponding observational results  and provides an accelerating universe.\\

\noindent
{\bf Acknowledgment:} The authors are very much thankful to the reviewer and editor for their constructive comments for the improvement of the paper.

\begin{table}[!h]
	\centering
	\caption{Results for Energy Conditions with $\xi=0$}
	\begin{tabular}{|c|c|l|}
		\hline
		\textbf{S.No.}& \textbf{Parameters} &\textbf{Results }\\
		\hline
		1 &$\gamma>0$, $k>0$ & $T>0$ and $\rho$ $> 0$, for all $t>0$, \\
		&& $\rho+p<0$, $\rho+3p<0$ and $\rho-|p|<0$, for all $t>0$\\\hline
		
		2 &$\gamma>0$, $k<0$ & $T$, $\rho$, $\rho+p$, $\rho+3p$ and $\rho-|p|$ are  negative, for all $t>0$ \\\hline
		3 &$\gamma>0$, $k=0$ & For $0<\gamma<0.3$, $T$, $\rho+p$ and $\rho-|p|$ are positive, for all $t>0$\\
		&&	$\rho<0$, $\rho+3p<0$, for all $t>0$\\
		&& For $\gamma\geq0.3$, $T>0$ and $\rho$ $> 0$, for all $t>0$\\
		&& $\rho+p$, $\rho+3p$ and $\rho-|p|$ are  negative, for all $t>0$ \\\hline
		
		4 &$\gamma<0$, $k>0$ & For $-1<\gamma<0$, $T$ and $\rho$ are positive, for all $t>0$ \\
		&& $\rho+p$, $\rho+3p$ and $\rho-|p|$ are   negative, for all $t>0$\\
		&& For $\gamma\leq -1$, $T$, $\rho$, $\rho+p$, $\rho+3p$ and $\rho-|p|$ are   negative, for all $t>0$\\
		\hline
		
		5 &$\gamma<0$, $k<0$ &$T$, $\rho$, $\rho+p$, $\rho+3p$ and $\rho-|p|$ are  negative, for all $t>0$\\\hline
		
		6 &$\gamma<0$, $k=0$ & For $-0.4<\gamma<0$, $T$, $\rho$, $\rho+p$, $\rho+3p$ and $\rho-|p|$ \\
		&&are complex, for all $t>0$\\
		&& For $\gamma\leq -0.4$, $T$, $\rho$, $\rho+p$,  $\rho+3p$ and $\rho-|p|$\\
		&& are negative, for all $t>0$\\\hline
		
		7 &$\gamma=0$, $k>0$ & $T$ and  $\rho$ are positive, for all $t>0$\\
		&&	$\rho+p>0$, for $t\in(0,0.2]\cup(5.6,\infty)$,\\
		&& $\rho+3p>0$, for $t\in(0,0.2]\cup(8.6,\infty)$, \\
		&& $\rho-|p|>0$, for $t\in(0,0.2]\cup(5.6,\infty)$\\\hline

		8 &$\gamma=0$, $k<0$ & $T>0$, $\rho$, $\rho+p$, $\rho+3p$ and $\rho-|p|$ are   negative, for all $t>0$\\\hline

		9 &$\gamma=0$, $k=0$ & $T>0$, $\rho$, $\rho+p$, $\rho+3p$ and $\rho-|p|$ are   complex, for all $t>0$\\\hline
		
	\end{tabular}
\end{table}

\begin{table}[!h]
	\centering
	\caption{Results for Energy Conditions with $\xi<0$}
	\begin{tabular}{|c|c|l|}
		\hline
		
		\textbf{S.No.}& \textbf{Parameters} &\textbf{Results }\\
		\hline
		1 &$\gamma>0$, $k>0$ & For $\xi<-6$, $T$, $\rho$, $\rho+p$, $\rho+3p$ and $\rho-|p|$ \\
		&&are negative or complex, for all $t>0$\\
		&& For $\xi\geq-6$, $T>0$ $\forall t>0$ and $\rho$ $> 0$, for $t\in(0,.0.195)\cup(0.24,\infty)$, \\
		&&$\rho+p>0$, for $t\in(0,0.01]\cup(0.37,\infty)$,\\
		&& $\rho+3p>0$, for $t\in(0,0.01]\cup(1.1,\infty)$, \\
		&& $\rho-|p|>0$, for $t\in(0,0.02]\cup(0.37,\infty)$\\\hline
		
		2 &$\gamma>0$, $k<0$ & For $-54<\xi<-1$, $T$, $\rho$, $\rho+p$, $\rho+3p$ and $\rho-|p|$ are positive, for all $t>0$\\
		&& For $\xi\leq-54$, $\xi\geq-1$, $T$, $\rho$, $\rho+p$, $\rho+3p$ and $\rho-|p|$ are  negative \\
		&&or complex, for all $t>0$\\\hline
		3 &$\gamma>0$, $k=0$&$T$, $\rho$, $\rho+p$, $\rho+3p$ and $\rho-|p|$ are  negative or complex, for all $t>0$\\\hline
		
		4 &$\gamma<0$, $k>0$ & For $\xi\leq-54$ or $\xi\geq -1.3$, $T$, $\rho$, $\rho+p$, $\rho+3p$ and $\rho-|p|$ \\
		&&are negative or complex, for all $t>0$\\
		&& For $-54<\xi<-1.3$, $T$, $\rho$, $\rho+p$,  $\rho+3p$ and $\rho-|p|$\\
		&& are positive, for all $t>0$\\\hline
		
		5 &$\gamma<0$, $k<0$ &$T$, $\rho$, $\rho+p$, $\rho+3p$ and $\rho-|p|$ are   complex, for all $t>0$\\\hline
		
		6 &$\gamma<0$, $k=0$ & For $\xi\leq-8.9$ or $\xi\geq -1$, $T$, $\rho$, $\rho+p$, $\rho+3p$ and $\rho-|p|$ \\
		&&are negative or complex, for all $t>0$\\
		&& For $-8.9<\xi<-1$, $T$, $\rho$, $\rho+p$,  $\rho+3p$ and $\rho-|p|$\\
		&& are positive, for all $t>0$\\\hline
		
		7 &$\gamma=0$, $k>0$ & For $\xi\leq-54$ or $\xi\geq -1$, $T$, $\rho$, $\rho+p$, $\rho+3p$ and $\rho-|p|$ \\
		&&are negative or complex, for all $t>0$\\
		&& For $-54<\xi<-1$, $T$, $\rho$, $\rho+p$,  $\rho+3p$ and $\rho-|p|$\\
		&& are positive, for all $t>0$\\\hline
		
		8 &$\gamma=0$, $k<0$ &$T>0$, $\rho$, $\rho+p$, $\rho+3p$ and $\rho-|p|$ are   complex, for all $t>0$\\\hline
		
		9 &$\gamma=0$, $k=0$& For $\xi\leq-53$ or $\xi\geq 0$, $T$, $\rho$, $\rho+p$, $\rho+3p$ and $\rho-|p|$ \\
		&&are complex, for all $t>0$\\
		&& For $-53<\xi<0$, $T$, $\rho$, $\rho+p$,  $\rho+3p$ and $\rho-|p|$\\
		&& are positive, for all $t>0$\\\hline

	\end{tabular}
\end{table}

\begin{table}[!h]
	\centering
	\caption{Hubble Parameter Observational data}
	\vspace{.5cm}
	\begin{tabular}{|c|c|c|c|c|}
		\hline
		S.No.& $z$ & $H(z)$ & $\sigma_i$&Reference\\
		\hline\hline
		1 & .090 & 69 & 12 & \cite{Jimenez1}\\\hline
		2 & .17 & 83 & 8 & \cite{Simon}\\\hline		
		3 & .27 & 77 & 14 & \cite{Simon}\\\hline	
		4 & .4 & 95 & 17 & \cite{Simon}\\\hline	
		5 & .9 & 117 & 23 & \cite{Simon}\\\hline
		6 & 1.3 & 168 & 17 & \cite{Simon}\\\hline
		7 & 1.43 & 177 & 18 & \cite{Simon}\\\hline
		8 & 1.53 & 140 & 14 & \cite{Simon}\\\hline
		9 & 1.75 & 202 & 40 & \cite{Simon}\\\hline
		10 & .48 & 97 & 62 & \cite{Stern}\\\hline
		11 & .88 & 90 & 40 & \cite{Stern}\\\hline
		12 & .179 & 75 & 4 & \cite{Moresco}\\\hline
		13 & .199 & 75 & 5 & \cite{Moresco}\\\hline
		14 & .352 & 83 & 14 & \cite{Moresco}\\\hline
		15 & .593 & 104 & 13 & \cite{Moresco}\\\hline
		16 & .68 & 92 & 8 & \cite{Moresco}\\\hline
		17 & .781 & 105 & 12 & \cite{Moresco}\\\hline
		18 & .875 & 125 & 17 & \cite{Moresco}\\\hline
		19 & 1.037 & 154 & 20 & \cite{Moresco}\\\hline
		20 & .44 & 82.6 & 7.8 & \cite{Blake}\\\hline
		21 & .60 & 87.9 & 6.1 & \cite{Blake}\\\hline
		22 & .73 & 97.3 & 7 & \cite{Blake}\\\hline
		23 & .07 & 69 & 19.6 & \cite{Zhang}\\\hline
		24 & .12 & 68.6 & 26.2 & \cite{Zhang}\\\hline
		25 & .2 & 72.9 & 29.6 & \cite{Zhang}\\\hline
		26 & .28 & 88.8 & 36.6 & \cite{Zhang}\\\hline
		27 & 1.363 & 160 & 33.6 & \cite{Moresco1}\\\hline
		28 & 1.965 & 186.5 & 50.4 & \cite{Moresco1}\\\hline
		29 & 2.34 & 222 & 7 & \cite{Delubac}\\\hline
	\end{tabular}
\end{table}
\end{document}